# Avocado Buying Trends in the United States Using SAC

Authors: Velma Jones, Kendra Keyse, Alfredo Melgoza, Karen Perez, Tammy Qamar, Jason Villalpando, Jongwook Woo
California State University Los Angeles BUS5100-93: Introduction To Business Analytics
{vjones, kkeyse, amelgoz9, kperez141, tqamar, jvilla19, jwoo5}@calstatela.edu

**Abstract:** The purpose of our paper is to analyze the dataset from Hass Avocado Board (HAB). The data features historical data on avocado prices and sales volume in multiple cities, states, and regions of the United States, ranging from 2015 to 2020. The paper consists of a mapped and calculated statistical analysis of the data over an established period. Using the cloud visualization tool SAP Analytics Cloud (SAC) to import and clean the data, we will complete comprehensive investigations and employ class lecture information as needed. Then, we will present insights using visualization, and time-series analysis. Our research intends to reveal insights into consumer buying statistics, such as the most popular type of avocado, preferences of organic to conventional and seasonality trends. The research is relevant as health-conscious trends have become increasingly popular, and avocado purchases indicate this.

## 1. Introduction

This paper uses SAP Analytics Cloud to process and visualize the historical data of avocados in the United States. The dataset was retrieved from Kaggle and it mainly consists of information on average prices of conventional versus organic avocados, total sales volume per type, type of avocados preferred, and geographical points of sales across the country in the last 5 years. We have chosen this dataset as the avocado has become a popular food item as health trends increase throughout the United States, especially due to COVID-19.

## 2. Related Work

Due to the significance avocados have as a standalone fruit as well as an integral ingredient in several cuisines, the price of avocados has been examined by numerous agriculture focused publications. The majority of these publications utilize the Hass Avocado Board's data whose mission is to collect data in order to determine trends and expected demand for avocados in the United States [1].

The first study that utilized the dataset compiled by the Hass Avocado Board is available on data centric division of the online publication website called Medium. It was conducted by George Washington University and focused on the correlation between price and volume sold for conventional and organic avocados [2]. This study chose to display its' findings through the use of scatter and group plots, and concluded conventional avocados have a moderate negative correlation between price and total units sold while organic avocados did not have any noticeable correlation [2]. Whereas our analysis, other than the analysis on price and volume sold, extends the trend analysis to the three common avocado sizes purchased throughout the United States using numerous comparison bar, bubble, and pie charts. Lastly, our study utilized a time series model with a forecast to create an interactive element to our study and allows any viewer to examine the total volume sold at a particular date.

The second study that utilized the same data was from an online publication called Agronometrics in Charts, which discusses an agricultural commodity on a weekly basis and the factors in the market that affect it. In June 2020 the installment selected avocados, used the same dataset in order to analyze pricing, and presented its' findings through multiple line charts [3]. The study tracked the changes in price over 2020 and attributed price anomalies to events such as the Super Bowl promotions aired at the beginning of the year [3]. As well, during mid-2018 and 2019 avocado supply volumes shifted from higher than average to normal which translated to higher prices due to an anticipation of a decrease in supply, whereas 2020 saw price stabilization around the same time [3]. In comparison to our work, we were able to extend our analysis to the end of 2020 and include a regression analysis that highlights influencers who currently and can potentially impact avocado sales; followed by the predictive element as to what we can expect the average price to be moving forward.

## 3. Specifications

The dataset was retrieved from Kaggle [4], which is an online community where data sets can be published and examined. The data has been updated on a consistent basis with data from the Hass Avocado Board to be inclusive of 2015 to 2020. From this statement we can infer the data set is straightforward enough to constantly integrate the current year's data. Therefore, a historical analysis can be conducted with avocado prices and total volume sold. The dataset consists of actual scan data from retailers' cash registers based on actual retail sales of Hass avocados as well as "multi-outlet reporting includes an aggregation of the following channels: grocery, mass, club, drug, dollar and military" [4]. The size of the data set is 3.37 MB and the following table indicates the data specifications within the total avocado sizes purchased.

Table 1. Data Specifications – Avocado Size

| Data Set | Size (Total 3.37 MB) |
|---|---|
| 4046 – Small/Medium Hass Avocado | 1688 KB |

| 4225 – Large Hass Avocado | 1562 KB |
|---|---|
| 4770 – Extra Large Hass Avocado | 120 KB |

## 4. Implementation Flowchart

The raw dataset downloaded from Kaggle comprised of avocado prices and sales volume from multiple cities, states, and regions of the USA. The data was scraped from the Hass Avocado Board and posted on Kaggle. The process of data manipulation is shown in the flowchart below. The dataset being used was provided in a single CSV file. The data file was uploaded to SAP Analytics cloud which was used to clean and create a story with models. The models were then exported to be use in a PowerPoint.

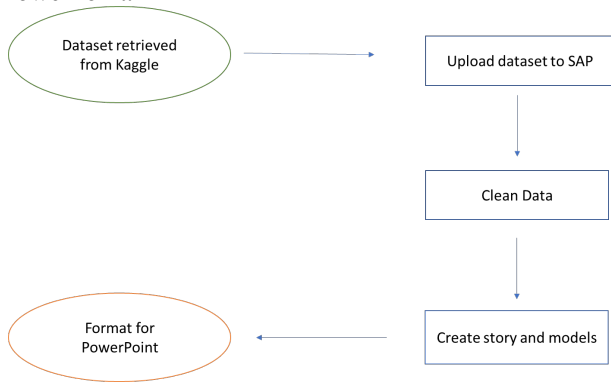

Figure 1. Implementation Flowchart

## 5. Data Cleaning

The CSV file was uploaded onto SAP Analytics cloud as a datasheet. From there, the first step in cleaning the data was to remove "Total US". This portion of the data was removed as it was skewing the results when used in certain models. Second the columns labeled "average price" and "year" were changed from measures to dimensions to allow for further manipulation. The dataset was then ready for use in to create a story in SAP.

## 6. Analysis and Visualization

After data cleaning and determining what data would best describe our analysis, a story and predictive model were created in SAP Analytics Cloud that provided a visual representation of the average price per year/type, total volume per type/location, a time series, and a predictive model for price/volume.

### 6.1 Average Price Per Type, Year

The first visualization (Figure 2), a bar chart, was created in SAP Analytics Cloud and shows the difference in average price per type of avocado (conventional/organic) during the timeline of years 2015-2020. The bar chart is clearly labeled and shows that the average price of organic avocados is generally always higher than conventional avocados.

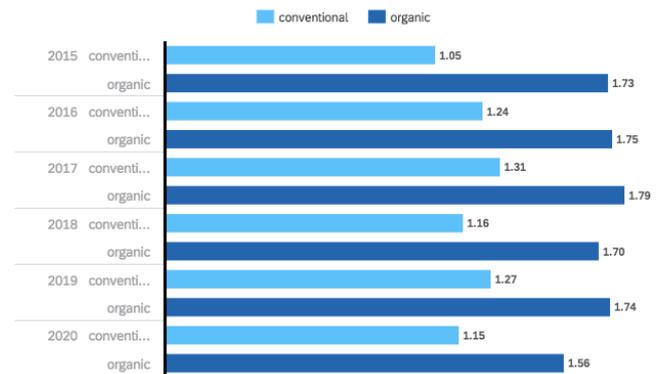

Figure 2. Average Price Per Type, Year of Avocados

### 6.2 Average Price Per Type, Conventional & Organic

The pie chart (Figure 3), which was also created in SAP Analytics Cloud, illustrates the average price per type of conventional and organic avocados. Nearly 58% of organic avocado sales averaged $1.80 per avocado and roughly 42% of conventional avocados averaged $1.30 per avocado.

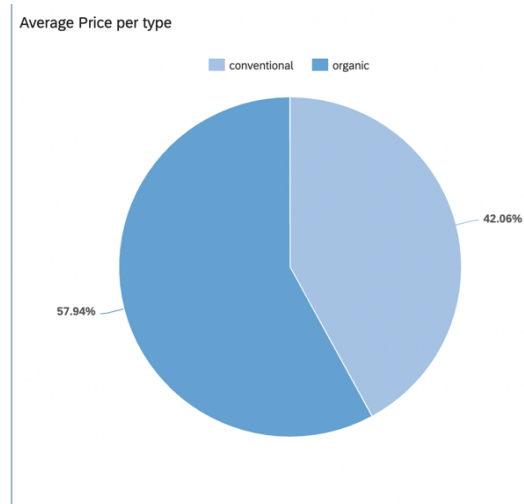

Figure 3. Average Price Per Type of Avocado

### 6.3 Total Volume Per Type & Year

The bar chart below (Figure 4) shows the historical data from years 2015-2020 of the total volume per type and year of avocados. We can see that the total volume output of conventional avocados far outnumbers the total volume output of organic avocados. We can also see that the total volume output of both conventional and organic avocados has grown steadily year after year.

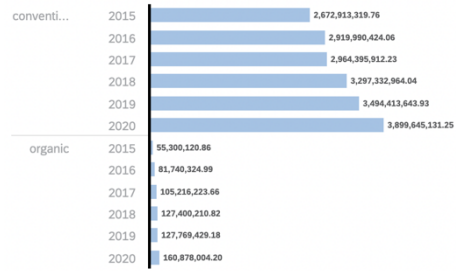

Figure 4. Total Volume Per Type & Year of Avocado

### 6.4 Time Series Analysis & Forecast of Avocado Size and Total Volume Purchases

The graph below (Figure 5) shows the time series analysis and forecast of avocado size and total volume purchased over the course of one year. 4046 represents small/medium Hass avocados, 4225 represents large Hass avocados, and 4770 represents extra-large Hass avocados. The graph clearly shows a gradual increase in total volume as the year progressed.

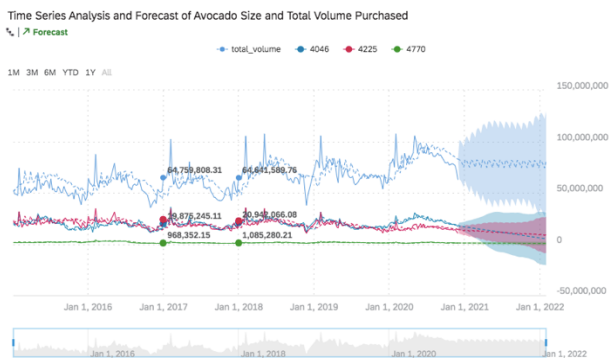

Figure 5. Time Series Analysis and Forecast of Avocado Size and Total Volume Purchases

### 6.5 Total Volume Purchased By Geography

The Figure 6 geo-map highlights the volume of avocados purchased by geographical location where the bubble color denotes the average price and the bubble size refers to total volume. California is the highest-ranking state at over 1.9 billion purchases, with the United States' western region leading overall at 2.15 billion. As the highest-ranking city, Los Angeles contributes significantly with over 959 million purchases. California's reputation as a health-conscious state and the avocado's reputation as a healthy food are likely contributing factors. Avocado demand has also improved due to the pandemic, with more health-conscious consumers eating at home [5]. Syracuse, NY, and Boise, ID are the lowest two cities on the chart with approximately 24.1 million and 30.3 million, respectively.

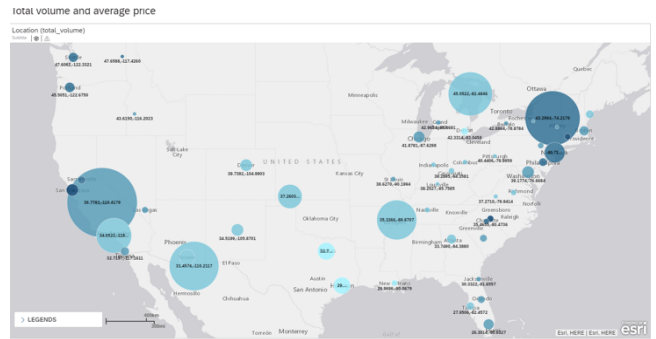

Figure 6. Total Volume Purchased By Geography Geo-Map

### 6.6 Regression Analysis

Using the Hass Avocado Board dataset, we created a predictive price model using regression analysis. The average price of avocados is used as the predictive goal. The root mean square error (RMSE) came to .12, where the closer to zero it is, the better the model due to fewer errors (Figure 7). The low score showcases how minimal the errors are in the model when comparing predicted vs. actual data. At the same time, Prediction Confidence is 98.29% (Figure 7). Prediction Confidence measures if the predictive model can make the predictions with the same reliability when new cases arrive; 100% is ideal.

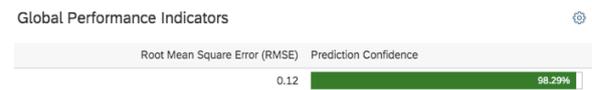

Figure 7. RMSE .12 & Prediction Confidence 98.29%

Figure 8 depicts Predicted vs. Actual chart data while showing the avocado prices. One of the Validation - Actual prices is $2.05 while the Predicted Value is $2.02. The regression analysis shows a Validation - Error Max of $2.24 and a Validation - Error Min of $1.86. A Perfect Model would have shown $2.03, so our model's $2.02 prediction is exceptionally close.

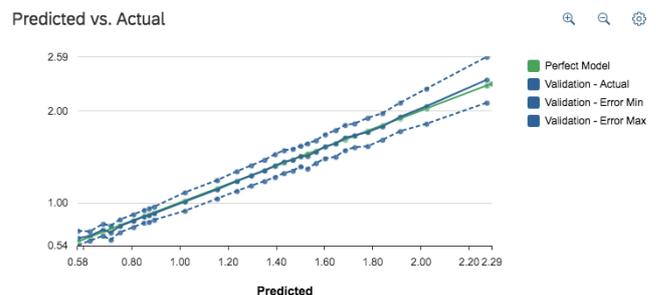

Figure 8. Predicted Avocado Prices vs. Actual Avocado Prices

The following Figure 9 highlights the influencers that contributed to this analysis. From highest to lowest are: which are the type of avocado (19.89%), geography (12.5%), year (12.15%), total purchase volume (11.76%), and date DOY [day-of-the-year] (10.27%):

| Influencer Contributions | |
|---|---|
| Influencer | Contribution |
| type | 19.89% |
| geography | 12.50% |
| year | 12.15% |
| total_volume | 11.76% |
| date_DoY | 10.27% |

Figure 9. Influencer Contributions

## 7. Conclusion

Finally, summing up all the above work we can conclude the following:
  i. The price of organic avocados is on average 35-40% higher than conventional avocados.
  ii. The sales volume of conventional avocados per year is on average 30 times bigger than that of the organic avocado sales
  iii. Seasonality trends reveal that the highest point of sales take place in early February as well as the first week of May
  iv. Avocado type, avocado price, and geographical location play a role in avocado buying behaviors